%% file: prd698.tex
\documentclass[twocolumn,showpacs,aps,prl,superscriptaddress]{revtex4}

\usepackage{graphicx}
\usepackage{dcolumn}
\usepackage{amsmath}
\usepackage{multirow}
\usepackage{amssymb}
\input babarsym

\def\bKForg {\rm {$B\ra K_2^*(1430) \gamma $}}
\def\bKForgn {\rm {$B^0\ra K_2^{*}(1430)^0 \gamma $}}
\def\bKForgc {\rm {$B^{+}\ra K_2^{*}(1430)^+ \gamma $}}

\newcommand{\bKThrg}{\rm $B\ra K^*(1410) \gamma $}

\def\cth	{\mbox{$|\cos{\theta_H}|\ $}}

\def\etal{{\it et al.}}
\newcommand{\BABARPubYear}    {04}
\newcommand{\BABARPubNumber}  {022}

\newcommand{\SLACPubNumber} {10104}

\def\figurebox#1#2#3{%
    \def\arg{#3}%
    \ifx\arg\empty
    {\hfill\vbox{\hsize#2\hrule\hbox to #2{\vrule\hfill\vbox to #1{\hsize#2\vfill}\vrule}\hrule}\hfill}%
    \else
    {\hfill\epsfbox{#3}\hfill}%
    \fi}

\begin{document}


\begin{flushleft}
\babar-PUB-\BABARPubYear/\BABARPubNumber\\
SLAC-PUB-\SLACPubNumber\\
\end{flushleft}

\title{
{\large \bf \boldmath
Measurement of the $B^{0}\ra K_2^{*}(1430)^0 \gamma \ $ and $B^{+}\ra K_2^{*}(1430)^+ \gamma \ $branching fractions} 
}

\input authors_jun2004.tex

\date{\today}

\begin{abstract}
We have investigated the exclusive, radiative $B$ meson decays to $K_2^*(1430)$ in $89\times 10^6$ \BB  events with the \babar$\ $detector at the PEP-II storage ring. We measure the branching fractions \BR (\bKForgn) $\ = (1.22\pm0.25\pm0.10)\times10^{-5}$ and \BR (\bKForgc)$\ = (1.45\pm0.40\pm0.15)\times10^{-5}$, where the first error is statistical and the second systematic. In addition, we measure the \CP-violating asymmetry ${\cal A}_{CP}$[\bKForgn]$\ = -0.08 \pm 0.15 \pm 0.01$.
\end{abstract}

\maketitle

In the Standard Model (SM), flavor-changing neutral currents (FCNC) are forbidden at the tree level. For example, there is no direct coupling between the $b$ quark and the $s$ or $d$ quarks. Effective FCNC are induced by loop (or ``penguin'') diagrams, where a quark emits and reabsorbs a $W$ thus changing flavor twice.
    
The discovery of $B\rightarrow K^*(892)\gamma$ decay~\cite{CLEOresult} verified the existence of penguin processes. The same publication also reported evidence for \bKForg, later confirmed by the BELLE collaboration~\cite{BELLEr}. Detailed knowledge about the decays to resonant modes with masses higher than $K^*(892)$, such as the $B \ra K_2^{*}(1430) \gamma$ decay, will provide a better understanding of the inclusive $b\ra s\gamma$ branching fraction in terms of the sum over exclusive modes~\cite{pseudo}. This is important because the comparisons between the inclusive theoretical and experimental rates place strong constraints on physics beyond the SM~\cite{Model}. The measurement of the $CP$ asymmetry, defined as ${\cal A}_{CP}=\frac{\Gamma\left(\Bbar\to\bar{f}\right)-\Gamma\left(B\to f\right)}{\Gamma\left(\Bbar\to\bar{f}\right)+\Gamma\left(B\to f\right)}$, places a further stringent test on the SM, because the theoretical uncertainty in the non-perturbative hadronic effects cancels~\cite{Theo}.

This study is based on 81 \invfb of data collected at the \FourS resonance (``on-resonance'') with the \babar\ detector at the \pep2\ asymmetric \ep(3.1\gev)-\en(9.0\gev) storage ring, corresponding to $89\times 10^6$ \BB pairs. We have also collected a data sample of $10$\invfb at 40\mev below the \FourS energy (``off-resonance'').

The \babar\ detector is described in detail elsewhere~\cite{ref:babar}. Charged particle trajectories are measured by a five-layer double-sided silicon vertex tracker (SVT) and a 40-layer drift chamber. Photons and electrons are measured in the barrel and forward end-cap electromagnetic calorimeters, consisting of 6580 thallium-doped CsI crystals.

Charged particle identification is provided by the energy loss ($dE/dx$) in the tracking devices and by a ring-imaging Cherenkov detector (DIRC). A $K/\pi$ Cherenkov angle separation better than 4 standard deviations is achieved for charged tracks with momenta below 3\gevc. 

We use Monte Carlo (MC) simulations of events in the \babar$\ $ detector based on GEANT4~\cite{Geant} to optimize our selection criteria and to determine signal efficiencies. These simulations take into account variations of the detector conditions and beam backgrounds over the data-taking period.

The $K_2^*(1430)$ is reconstructed from three modes $K_2^{*}(1430)^0\ra K^+\pi^-$ and $K_2^{*}(1430)^+\ra K^+\pi^0, K^0\pi^+$. $K^0$ mesons are reconstructed from the decay $K^0_S\ra \pi^+\pi^-$. Here and throughout this paper the charge-conjugate decays are included implicitly unless otherwise stated.

A photon candidate is defined as a localized energy deposition well contained within the calorimeter acceptance, $-0.77<\cos\theta <0.96$, where $\theta$ is the polar angle with respect to the detector axis. It must have a lateral energy profile consistent with a photon shower and  must be separated by 25\cm from all other showers, both neutral and charged. To suppress photons from $\pi^0 (\eta)$ decays, we veto any photon candidate that combines with another photon of energy greater than $50\ (250)$ \mev to form a $\gamma\gamma$ invariant mass in the range $115 < M_{\gamma \gamma } < 155\ (508 < M_{\gamma \gamma } < 588)$\mevcc. 

The $\pi^0$ candidates are reconstructed from pairs of photons that have an energy above 50 \mev and an opening angle less than 36 degrees; the invariant mass of the two photons is required to be in the range $115 < M_{\gamma\gamma} < 150$\mevcc. The candidate's momentum is recalculated with a $\pi^0$ mass constraint to improve the energy resolution.

The $K^\pm$ and $\pi^\pm$ track candidates are required to be consistent with originating from the \epem interaction point (IP); this requirement rejects tracks from beam-material and beam-gas interactions.
A track is identified as a kaon if it passes through the DIRC radiators, and the detected Cherenkov photons are consistent in time and angle with a kaon of the measured track momentum. A charged pion is defined as a track that is not identified as a kaon or an electron, based on $dE/dx$ and the ratio of the track momentum to the associated shower energy in the CsI calorimeter.

The $K^0_S$ candidates are reconstructed from two oppositely charged tracks, having an invariant mass satisfying $489<M_{\pi^+\pi^-}<507$\mevcc. We require that the $K^0_S$ candidate form a vertex that is displaced from the IP by at least 0.2\cm and lie in a direction from the IP consistent with the $K^0_S$ momentum. 

The $K_2^*(1430)$ candidate is required to have a $K\pi$ invariant mass within $120\ (110)$ \mevcc of the known $K^{*0}_2$ $(K^{*+}_2)$ mass~\cite{ref:pdg2002}. For the $K^+\pi^-$ mode, we require that the two tracks are consistent with originating from a common vertex. 

The $B$ candidates are reconstructed by combining one $K_2^*(1430)$ and one $\gamma$ candidate. To isolate the $B$ meson signal, we use two kinematic variables. The first, \DeltaE, is defined as the difference between the reconstructed energy of the $B$ candidate and the beam energy, which is known to high precision. The second is the beam energy substituted mass (\mes), which is defined as $m_{\rm ES}^{\rm raw}=\sqrt{E_{\rm beam}^{2}-p_B^{2}}, {\rm where\ } E_{\rm beam}=\sqrt{s}/2,\ \vec{p}_{B}=\vec{p}_{K^*}+\vec{p}_{\gamma}$ with $\vec{p}_{K^*}$ and $\vec{p}_{\gamma}$ representing the momenta of the $K_2^{*}$ and the photon candidates. For signal events, \DeltaE and $m_{\rm ES}^{\rm raw}$ peak at zero and at the $B$ meson mass, $m_B$, respectively. For the modes containing a single photon candidate, namely $K^+\pi^-$ and $K^0_S\pi^+$, we adopt a technique~\cite{CLEOresult} that rescales the measured photon energy in the center-of-mass (CM) frame (denoted by asteroids) $E^*_{\gamma}$ with a factor $\kappa$, determined for each event, such that $E_{K^*}^* + \kappa E_{\gamma}^* = E_{\rm beam}$ in the rest frame of the \FourS; this improves the original \mes\ ($m_{\rm ES}^{\rm raw}$) resolution from 3.0 to 2.7\mevcc. We retain $B$ candidates with the invariant mass closest to the $K_2^*(1430)$ mass if we find multiple candidates with $|\DeltaE|<0.3$\gev and \mes$>5.2$\gevcc in the same event, which occurs in $3.1$, $6.3$, and $4.9\%$ of the events for the $K^+\pi^-$, $K_S^0\pi^+$ and $K^+\pi^0$ modes, respectively.
 
The background has two components, one of which includes combinatorial background from $B$ decays and continuum \qqbar production, where $q$ can be a $u$, $d$, $s$ or $c$ quark, with the high-energy photon originating from initial-state radiation (ISR) or from $\pi^0$ and $\eta$ decays. These backgrounds are non-peaking in \mes and \DeltaE.

The second background contribution is from other resonant $B\ra X_s\gamma$ modes, predominantly \bKThrg, and non-resonant $B\ra K\pi\gamma$ decays. We label these the ``peaking'' background, since these decays have \mes and \DeltaE distributions similar to the signal. In order to distinguish the \bKForg$\ $signal from the background decays, we examine the helicity-angle distributions. The helicity-angle $\theta_H$ is defined as the angle of the $K^+$ or $\KS$ in the rest frame of the $K_2^*(1430)$ with respect to the flight direction of the $K_2^*(1430)$, measured in the $B$ meson rest frame. These modes have different helicity-angle distributions: ${\sin}^2\theta_{H}{\cos}^2\theta_{H}$ for $K_2^*(1430)$, ${\sin}^2\theta_{H}$ for $K^*(1410)$ and primarily ${\sin}^2\theta_{H}$ for non-resonant decays assuming $J=1$ for the spin of the $K\pi$ system. The non-resonant decays may have higher angular momentum contributions but the lowest possible angular momentum state is dominant; therefore, the helicity-angle distribution for the non-resonant decay is assumed to be the same as that of the \bKThrg$\ $ decay. The systematic uncertainty associated with this modeling is studied and included in the measured branching fraction uncertainty. 

\begin{figure}[htb]
\begin{center}   
\begin{tabular}{cc} 
\mbox{\includegraphics[width=1.7in]{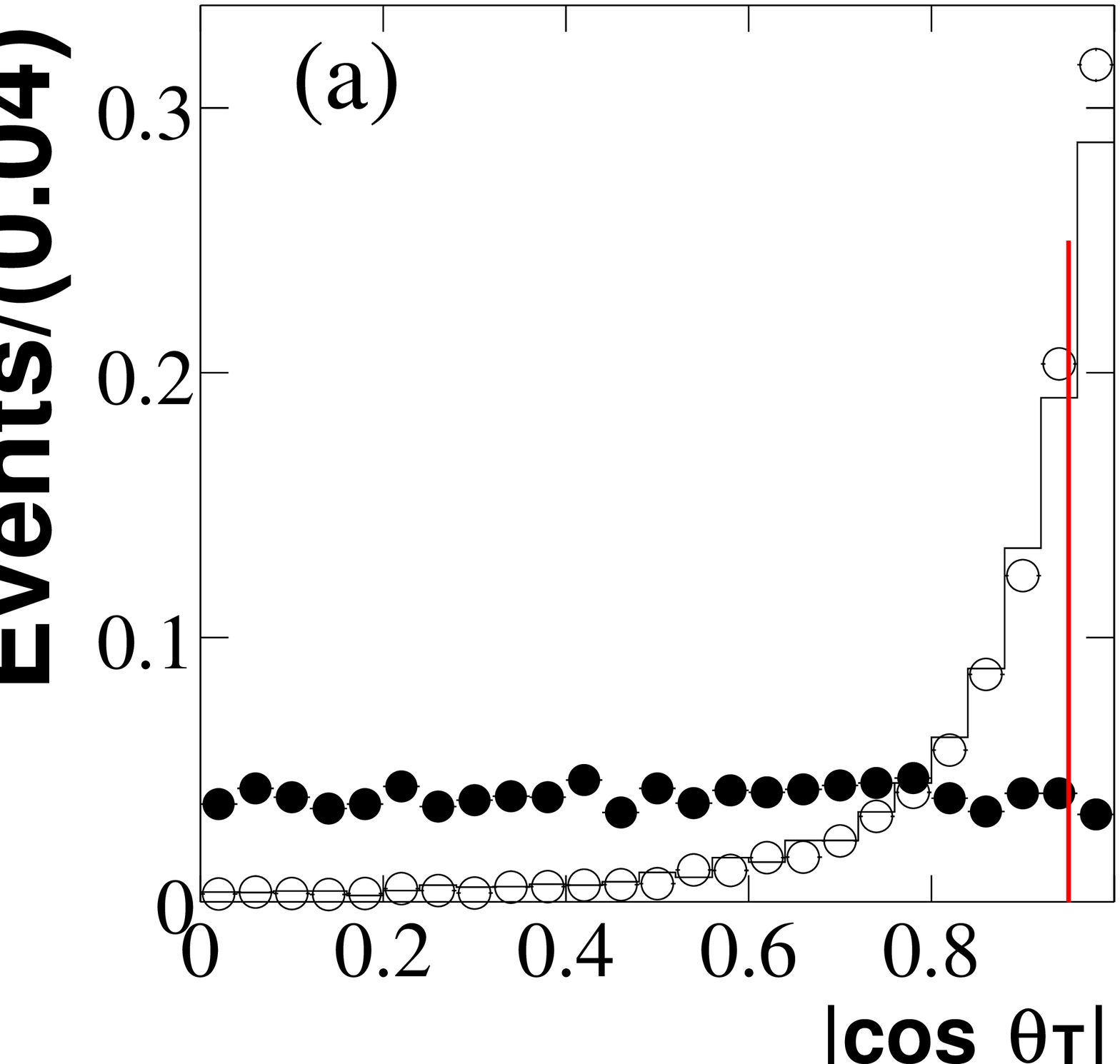}
\includegraphics[width=1.7in]{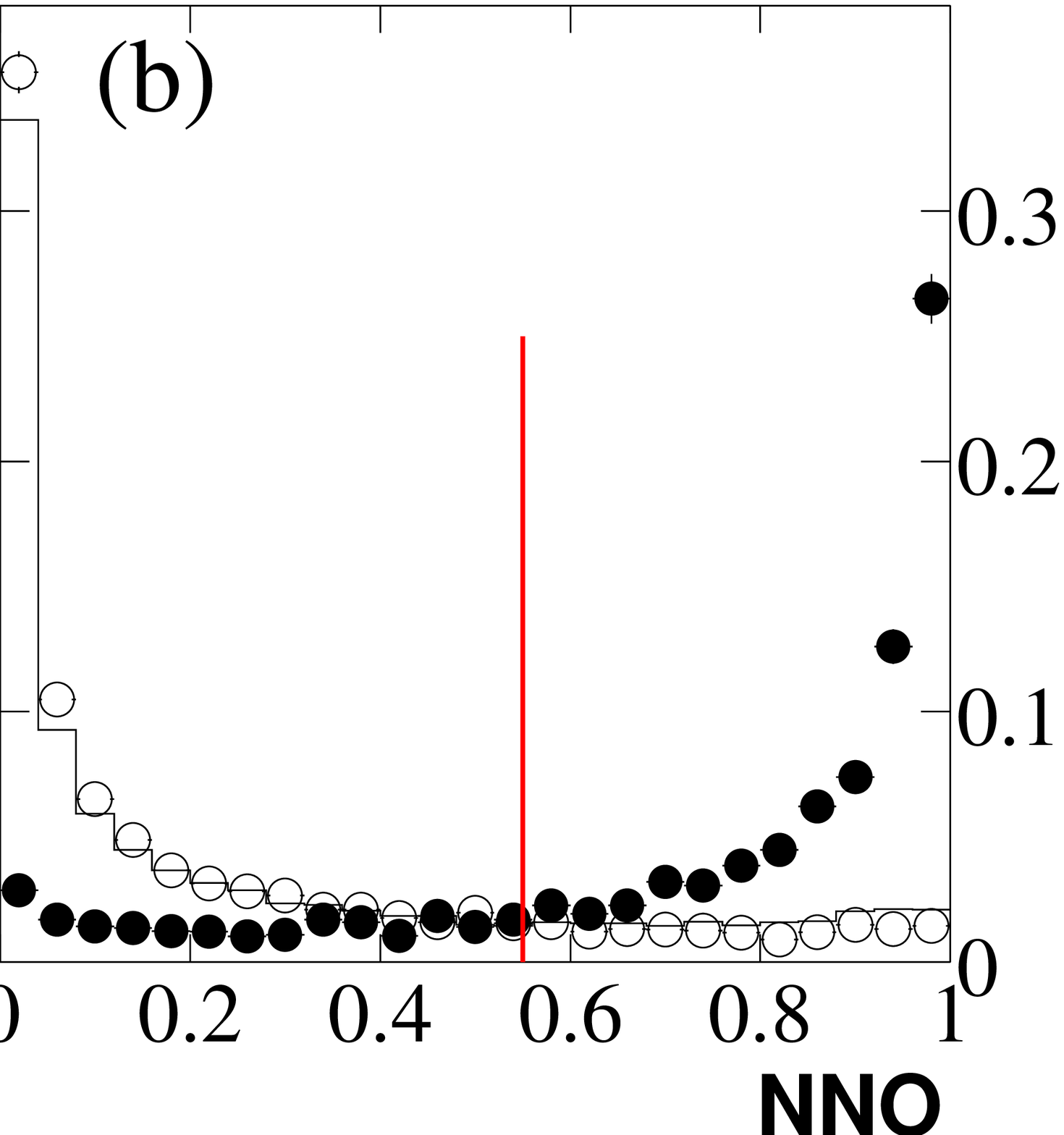}}
\end{tabular}
\end{center}    
\caption[fig:cosT]{ The cosine of thrust angle (a) and neural network output (b) distributions of the \bKForg$\ $MC simulation (filled circles), the off-resonance data (line), and the continuum background MC (open circles). The vertical line indicates the cut value.
\label{fig:cosT}}
\end{figure}

We exploit the difference in the event topology between signal and continuum background to reduce the continuum contribution. To remove radiative Bhabha and $\epem \ra\tau^+\tau^-$ events, we require that the ratio of second-to-zeroth order Fox-Wolfram moments~\cite{FoxWolf} of the event be less than 0.9. The distribution of the thrust angle $\theta_{\rm T}$, defined as the angle between the direction of the photon candidate and the thrust axis of the rest of the event in the CM frame, is shown in Fig.~\ref{fig:cosT}(a). The rest of the event includes all the particles not used in the reconstruction of the $B$ candidate.

We train a neural network~\cite{SNNSNET} with a combination of the thrust angle, the angle of the $B$ meson candidate's direction with respect to the beam axis, the scalar sum of CM momentum of the rest of the event~\cite{FisherCone} (binned with $10^{\circ}$ intervals ranging from parallel to anti-parallel relative to the photon momentum), sphericity, and the ratio of second-to-zeroth order Fox-Wolfram moments in the photon recoil system, which suppresses ISR background. The neural network improves background suppression significantly. The distribution of the neural network output (NNO) is shown in Fig.~\ref{fig:cosT}(b) for MC signal, MC continuum background and off-resonance data.

The cuts on thrust angle and neural network output have been optimized for the best statistical significance; an iterative method of optimization is used to minimize correlations. The optimized cuts are $|\cos\theta_{\rm T}| < 0.95$ and NNO$ > 0.55$, as indicated in Fig.~\ref{fig:cosT}.

The signal yields are extracted using a simultaneous maximum-likelihood fit of the \mes, \DeltaE and \cth distributions. The fit is performed independently for each of the decay modes considered here.

The signal \mes and \DeltaE distributions are well described by an asymmetric resolution function (``Crystal-Ball" function~\cite{Crystal}), having an approximately Gaussian core plus a long tail due to the energy leakage from the calorimeter for the photon candidates.
The peaking background is assumed to have the same \mes and \DeltaE distributions as the signal. The continuum background is parameterized empirically by an ARGUS function~\cite{BGpdf} for \mes and a linear function for \DeltaE.

The $\cos{\theta_H}$ distribution of the signal has been parameterized with ${\sin}^2{\theta_H}{\cos}^2{\theta_H}-\lambda({\cos}^4{\theta_H}-{\cos}^6{\theta_H})$, where $\lambda$ is a parameter determined from the Monte Carlo sample to account for the effect of the detector acceptance and efficiency. The \cth distribution of the ``non-peaking'' background is parameterized by a linear combination of exponential and constant components. 
 
\begin{figure}[htb]
\begin{center}   
\begin{tabular}{c} 
   \mbox{\includegraphics[width=3.3in]{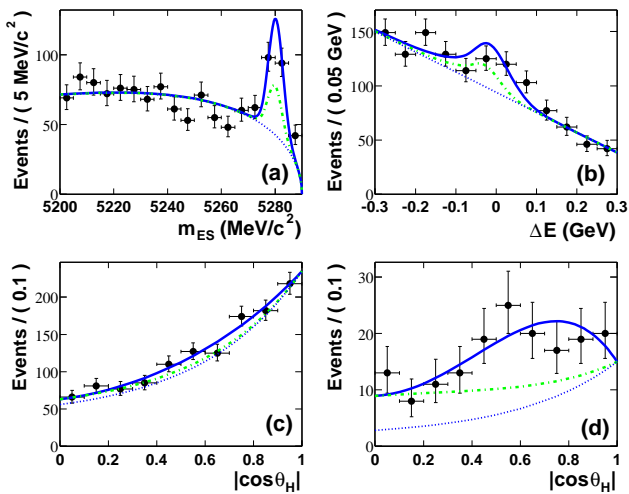}}
\end{tabular}
\end{center} 
\caption[fig:kpionr]{Distributions of (a) \mes, (b) \DeltaE, and (c) \cth for \bKForgn, $K_2^{*}(1430)^0\ra K^+\pi^-$ candidates in data, and (d) \cth in the signal region. The solid line shows the result of the fit to the data. The peaking (dashed-dotted line) and non-peaking (dotted line) background contributions are also shown.
\label{fig:kpionr}}
\end{figure}

\begin{figure}[htb]
\begin{center}   
\begin{tabular}{c} 
   \mbox{\includegraphics[width=3.3in]{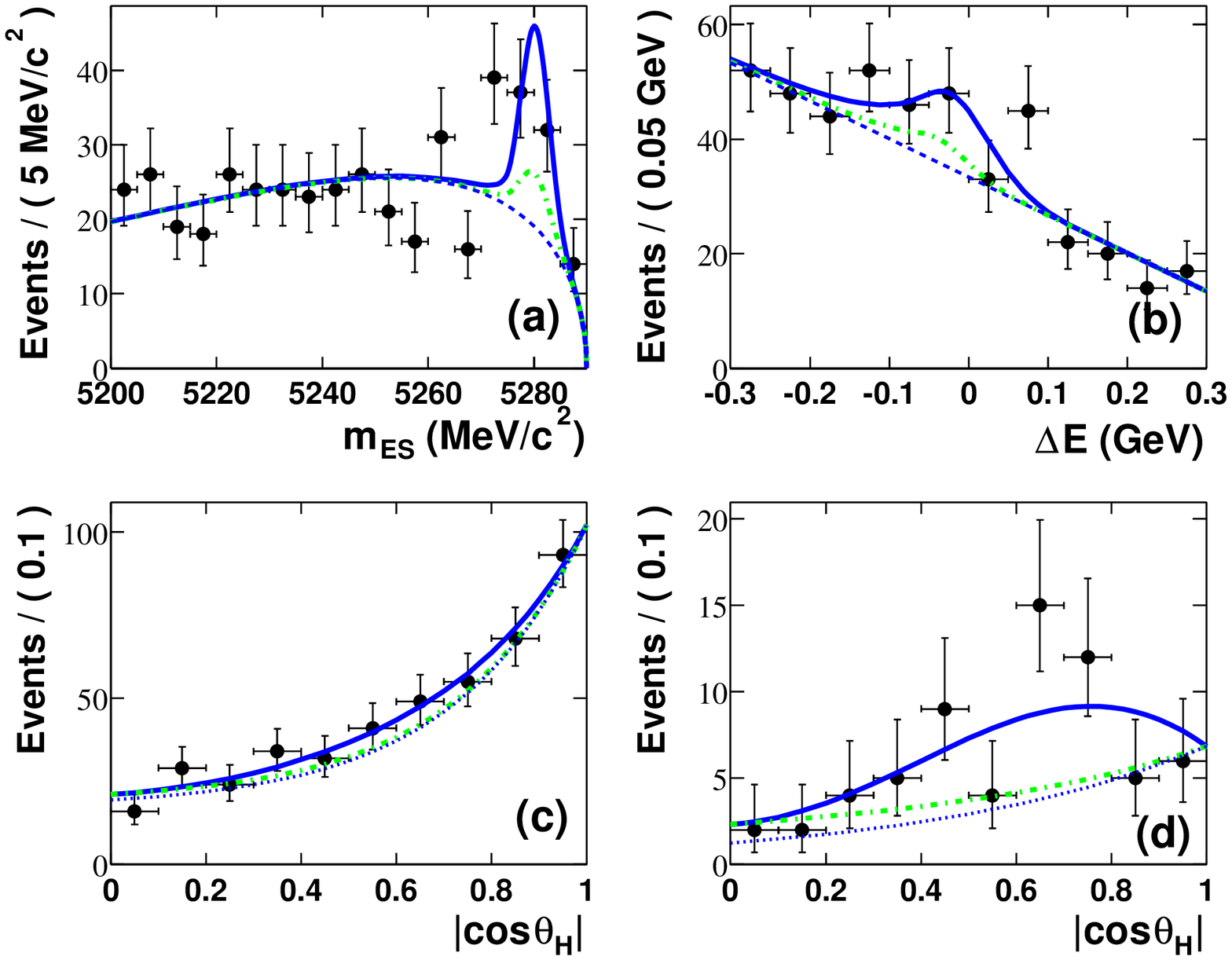}}\\
  \mbox{\includegraphics[width=3.3in]{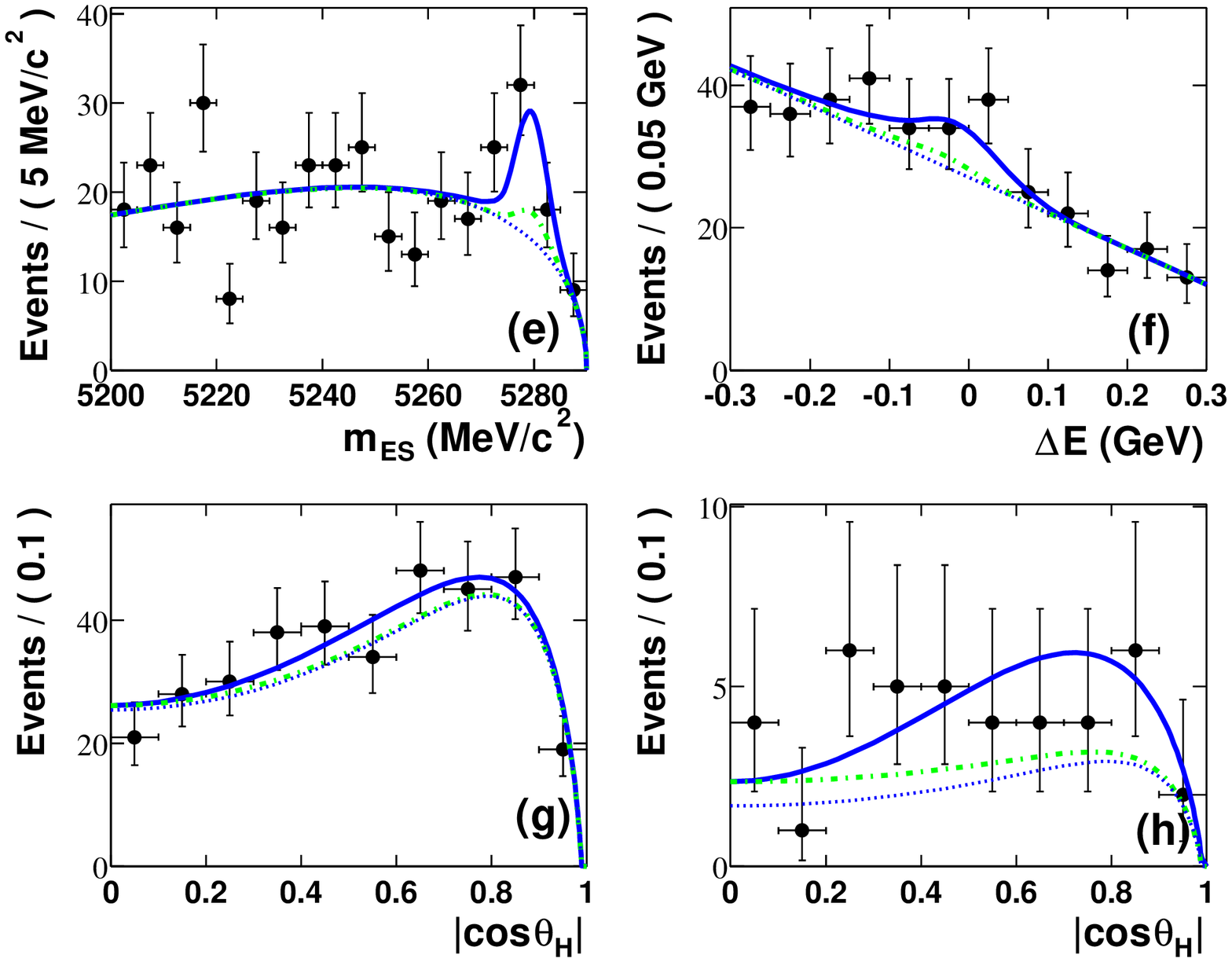}}
\end{tabular}
\end{center} 
\caption[fig:kspionr]{Distributions of (a) \mes, (b) \DeltaE, and (c) \cth for the \bKForgc, $K_2^{*}(1430)^+\ra K_S^0\pi^+$ candidates in data, and (d) \cth in the signal region. The solid line shows the result of the fit to the data. The peaking (dashed-dotted line) and non-peaking (dotted line) background contributions are also shown. The corresponding distributions for $K_2^{*}(1430)^+\ra K^+\pi^0$ candidates are shown in (e-h). 
\label{fig:kspionr}}
\end{figure}

Figures~\ref{fig:kpionr} and \ref{fig:kspionr} show the \mes, \DeltaE, and \cth distributions for the three modes in data; also shown are the \cth distributions of the candidates in the signal region, $-0.15< \DeltaE <0.10$\gev and $5.272<\mes<5.288$\gevcc. The signal as well as background yields are allowed to vary in the fit. All the non-peaking background parameters are determined by the fit. The signal and peaking-background helicity-angle, Crystal-Ball width, and shape parameters are fixed to the MC expectations. The means of the signal \mes and \DeltaE functions are fixed to the MC expectations, calibrated using $B\ra K^*(892)\gamma$ candidates from MC simulation and data, while the peaking-background means are allowed to float due to their complex composition. The signal yields are given in Table~\ref{tab:yields}. The signal significance has been evaluated from the change in the likelihood when the fit is repeated with the signal yield set to zero, including the systematic uncertainties, which are assumed to be normally distributed. The branching fractions are calculated with the assumption that the \FourS decays equally to neutral and charged $B$ meson pairs.

\begin{table}[btp]
\caption{The efficiency, fitted signal yield, significance, and measured branching fraction \BR (\bKForg) for each $K_2^{*}(1430)$ decay mode.}
\footnotesize 
\begin{tabular}{lcccc }      \hline \hline
 Mode & $\epsilon$ ($\%$) &  Signal  &Significance ($\sigma$)& $\cal{B}$($10^{-5}$) \\ \hline
  $K^+\pi^-$  & 6.4   & $69\pm14$  & 5.7 & $1.22\pm0.25\pm0.10$  \\ 
 $K^0_S\pi^+$ & 1.9 & $29\pm10$ & 3.1  & $1.69\pm0.59\pm0.16$   \\  
 $K^+\pi^0$ & 1.9 & $20\pm9$ & 2.2 & $1.23\pm0.55\pm0.15$ \\ \hline \hline
\end{tabular}
\label{tab:yields}
\end{table}

\begin{figure}[htb]
\begin{center}   
\begin{tabular}{c} 
   \mbox{\includegraphics[width=1.7in]{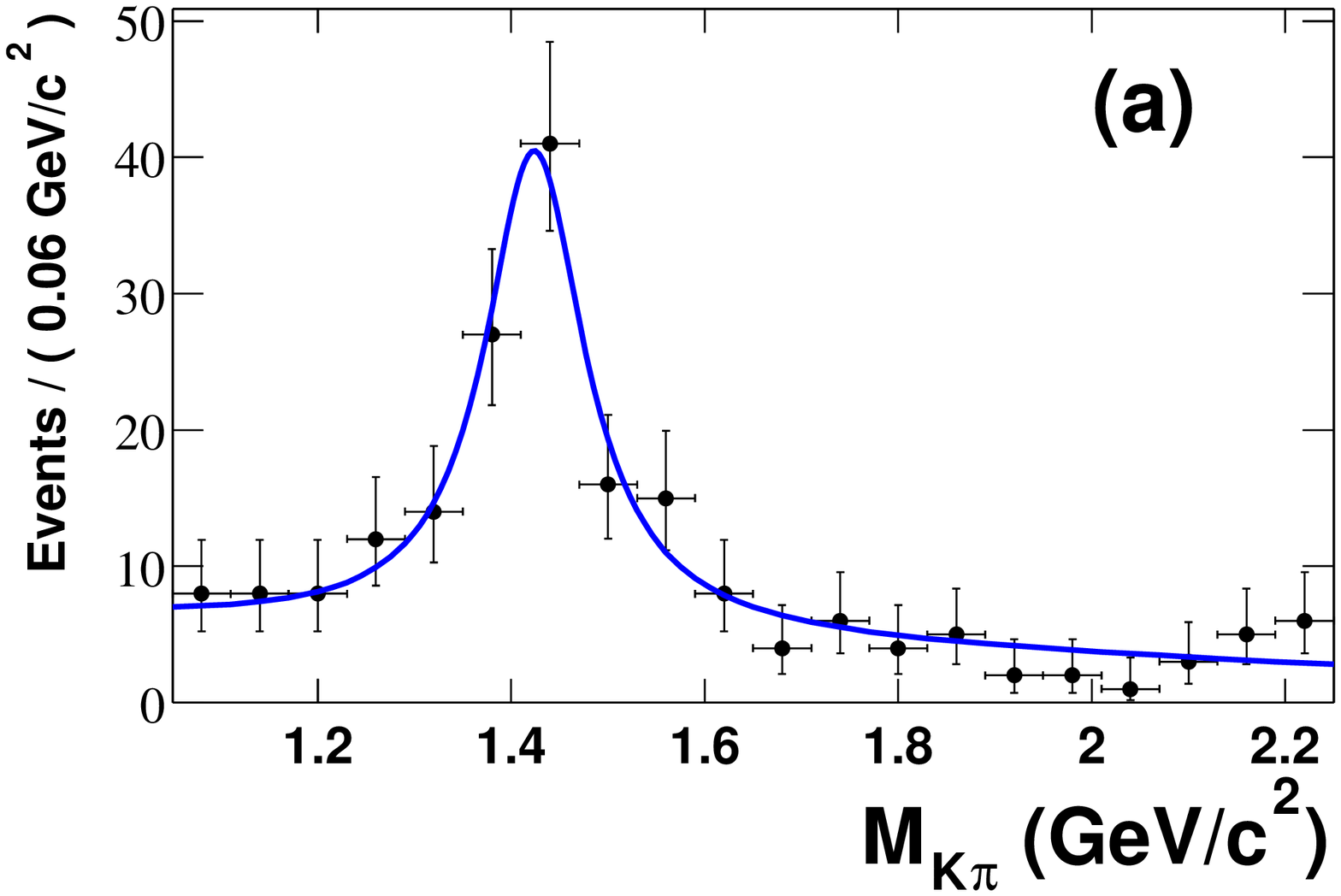}
	\includegraphics[width=1.7in]{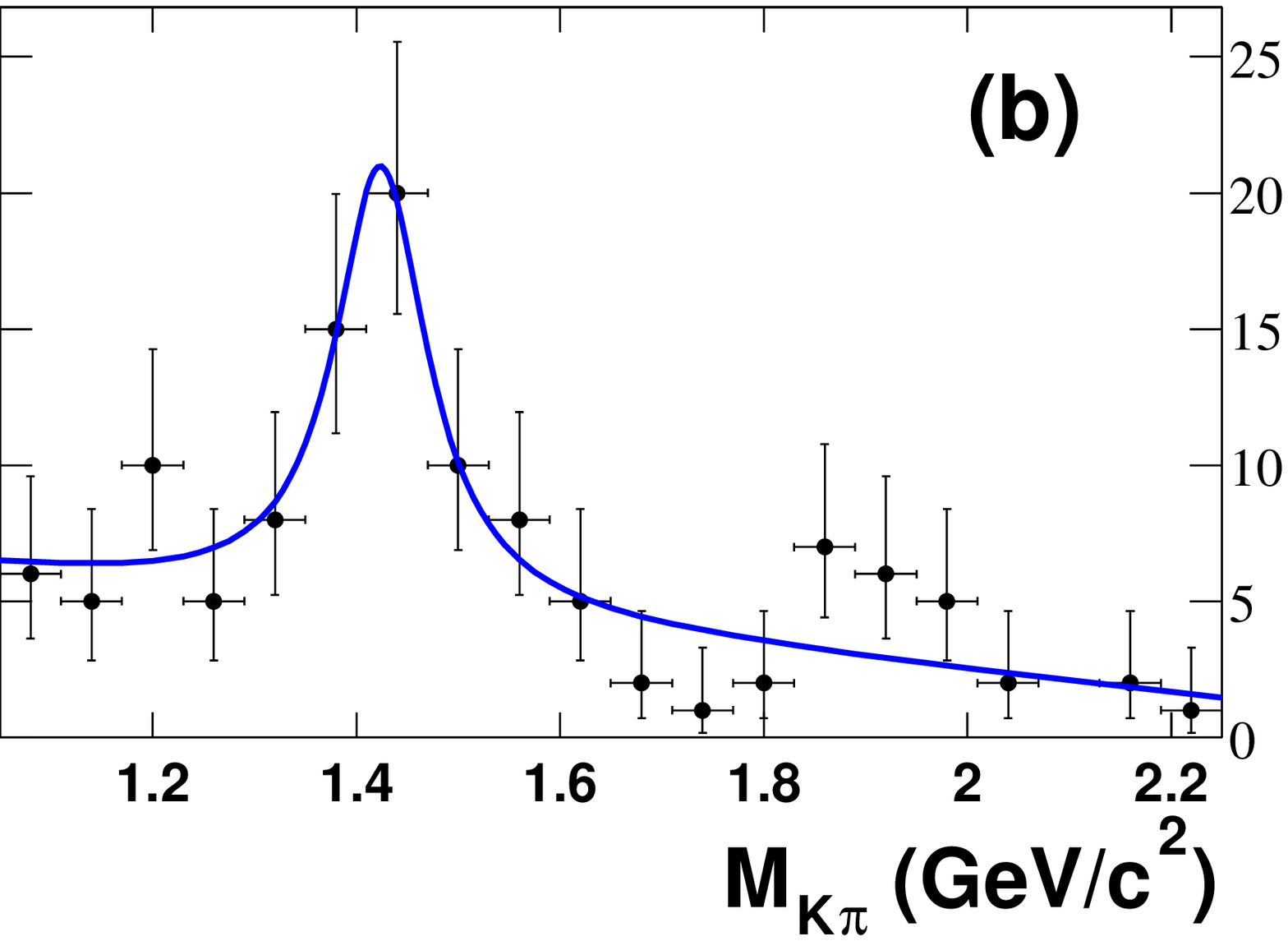}}
\end{tabular}
\end{center} 
\caption[fig:mksproject]{ (a) $K^+\pi^-$ and (b) $K^0_S\pi^+$ and $K^+\pi^0$ invariant mass distributions for the signal region (see text) after background subtraction.
\label{fig:mksproject}}
\end{figure}

Figure~\ref{fig:mksproject} shows the $K\pi$ invariant mass distribution where the cut on this quantity has been relaxed. The non-peaking background, estimated from the data outside the signal region, has been subtracted. The invariant mass is fit with a relativistic Breit-Wigner function plus a first-order polynomial. There is a clear enhancement around 1.4\gevcc in both the neutral and charged modes.

We use the kaon charge to tag the flavour and measure the direct \CP\ asymmetry in the decay rate as ${\cal A}_{CP}$[\bKForgn]$\ = -0.08\pm0.15$.

The systematic error on the branching fraction for each mode is shown in Table~\ref{tab:syst}. The total systematic uncertainty is computed as the sum in quadrature of the components. The yields returned from the fit are divided by the number of \BB\ events and corrected for the efficiency to obtain the branching fraction; therefore, the $1.1\%$ uncertainty on the $B$-counting measurement is included. The \DeltaE resolution is dominated by the photon energy resolution, which is determined from data using $\pi^0$ and $\eta$ meson decays with symmetric daughter photon energies. The deviation in the reconstructed $\eta$ mass from the nominal $\eta$ mass provides an estimate of the uncertainty in the measured single photon energy. The photon isolation and $\pi^0/\eta$ veto efficiency depend on the event multiplicity, and the effect is estimated by ``embedding'' MC-generated photons into both an exclusively reconstructed $B$ meson data sample and a generic $B$ meson MC sample. The photon and $\pi^0$ efficiency uncertainties are determined from a comparison of the efficiencies in data and MC for $\epem\ra\tautau$ events. The uncertainty in tracking efficiency is estimated from a sample of tracks well measured in the SVT. We estimate the uncertainties in the $K^0_S$ efficiency by comparing the data and MC distributions of the momentum and flight distance. The efficiency for kaon and pion identification in the DIRC is derived from a sample of the decays $D^{*+}\ra D^0\pi^+$, with $D^0\ra K^-\pi^+$. 

Because the variables used for the neural network training are mostly calculated through the information from the rest of the event, we use 3155 fully reconstructed $B\ra D\pi^-$ candidates in data, as well as simulated $B\ra D\pi^-$ events, as control samples. The pion in the $B\ra D\pi^-$ decay is treated like the photon in the \bKForg$\ $decay for the calculations of the event variables; the difference in the efficiency of the selection on the neural network output between data and MC is used as the systematic uncertainty. The systematic error also includes the uncertainty in the invariant mass and width of $K_2^*(1430)$ and its sub-mode branching fractions~\cite{ref:pdg2002}.

We estimate the systematic error due to the fitting procedures as follows. For the shape parameters of \mes, \DeltaE, and \cth distributions, we vary the parameters in the fit within their errors from the MC expectations. We also test the validity of the peaking-background \cth probability density function (PDF) by mixing up to $20\%$ $J=2$ components and generating MC samples with different PDF parameterizations. We use the largest deviation in these tests as the systematic error of the signal yield. There is also a systematic error associated with the limited statistics of the signal MC sample.

\begingroup
\begin{table}[btp]
\caption{Fractional systematic uncertainties ($\%$) in the measurement of \BR(\bKForg).}
\footnotesize
\begin{center}
\begin{ruledtabular}
\setlength{\extrarowheight}{1.5pt}
\begin{tabular}{lccc}  
Uncertainty &  $K^+ \pi^-$ & $\KS \pi^+$ & $K^+ \piz$ \\ \hline 

\BB\ events counting                    & 1.1      &1.1    & 1.1  \\
Photon and $\pi^0$ detection efficiency  & 2.5       &2.5    &7.5  \\
Photon energy scale                      & 1.0       & 1.0   & 1.0 \\
Photon energy resolution                 & 2.5       & 2.5   & 2.5 \\
Photon isolation                      & 2.0       & 2.0   & 2.0 \\
$\pi^0/\eta$ veto                        & 1.0       & 1.0   & 1.0 \\
Tracking efficiency                     & 1.6       & 0.8   & 0.8 \\
Kaon identification efficiency          & 1.0        & --   & 1.0 \\
Pion identification efficiency           & 0.6        & 0.6    & -- \\
$K^0_S$ efficiency                       & --       & 3.0   & --\\
Sub-mode branching fraction                    & 2.4      & 2.4   & 2.4 \\
$K_2^*(1430)$ mass/width                        & 1.6      & 1.0   & 1.1   \\
Signal PDF parameters                    & 3.9      & 5.8    & 6.3   \\
Background modeling                   & 2.6     & 2.9   & 2.9 \\
Peaking-background modeling             & 3.5      & 4.9   & 4.8   \\
MC statistics                            & 2.5      & 3.2   & 3.2  \\ \hline
Total                                    & 8.4      & 10.2   & 12.6  \\ 
\end{tabular}
\end{ruledtabular}
\end{center}
\label{tab:syst}
\end{table}
\endgroup

The particle-antiparticle asymmetry in the detector response, which includes $0.35\%$ uncertainty for the tracking efficiency and $1.0\%$ uncertainty for the charged particle identification, predominantly contributes to the systematic uncertainty of the ${\cal A}_{CP}$ measurement. The uncertainty in the estimate of nuclear interaction asymmetry, which arises from the different interaction probabilities of $K^+$ and $K^-$ and of $\pi^+$ and $\pi^-$, is $0.20\%$~\cite{NuInAsm}. The total ${\cal A}_{CP}$ systematic uncertainty is $1.1\%$.

We have presented a measurement of the branching fraction for \bKForgn$\ $of $(1.22\pm0.25\pm0.10)\times10^{-5}$, which has a $5.7\sigma$ statistical significance; this is in agreement with, but more precise than, previous experimental results. We observe a signal with a statistical significance of $3.8\sigma$ for \bKForgc$\ $ and measure the branching fraction to be $(1.45\pm0.40\pm0.15)\times10^{-5}$, by combining the results from $K_S^0\pi^+$ and $K^+\pi^0$ modes. Both results agree with the theoretical predictions based on a relativistic form-factor model~\cite{Theo}. The ${\cal A}_{CP}$ is measured to be $-0.08\pm 0.15 \pm 0.01$, thus no evidence of direct \CP\ violation is observed.

\section{Acknowledgments}
\label{sec:Acknowledgments}

\input acknowledgements

\end{document}

%% file: authors_jun2004.tex
%
\author{B.~Aubert}
\author{R.~Barate}
\author{D.~Boutigny}
\author{F.~Couderc}
\author{J.-M.~Gaillard}
\author{A.~Hicheur}
\author{Y.~Karyotakis}
\author{J.~P.~Lees}
\author{V.~Tisserand}
\author{A.~Zghiche}
\affiliation{Laboratoire de Physique des Particules, F-74941 Annecy-le-Vieux, France }
% ...
\collaboration{The \babar\ Collaboration}
\noaffiliation

%% file: acknowledgements.tex
We are grateful for the 
extraordinary contributions of our \pep2\ colleagues in
achieving the excellent luminosity and machine conditions
that have made this work possible.
The success of this project also relies critically on the 
expertise and dedication of the computing organizations that 
support \babar.
The collaborating institutions wish to thank 
SLAC for its support and the kind hospitality extended to them. 
This work is supported by the
US Department of Energy
and National Science Foundation, the
Natural Sciences and Engineering Research Council (Canada),
Institute of High Energy Physics (China), the
Commissariat \`a l'Energie Atomique and
Institut National de Physique Nucl\'eaire et de Physique des Particules
(France), the
Bundesministerium f\"ur Bildung und Forschung and
Deutsche Forschungsgemeinschaft
(Germany), the
Istituto Nazionale di Fisica Nucleare (Italy),
the Foundation for Fundamental Research on Matter (The Netherlands),
the Research Council of Norway, the
Ministry of Science and Technology of the Russian Federation, and the
Particle Physics and Astronomy Research Council (United Kingdom). 
Individuals have received support from 
CONACyT (Mexico),
the A. P. Sloan Foundation, 
the Research Corporation,
and the Alexander von Humboldt Foundation.